\documentclass[compsoc]{IEEEtran}
\usepackage{subcaption}
\usepackage{algorithm}
\usepackage{algorithmicx}
\usepackage{algpseudocode}
\usepackage{booktabs}

\usepackage[bottom]{footmisc}
\usepackage{amssymb}
\usepackage[nolist]{acronym}
\usepackage{amsmath}
\usepackage{amsfonts}
\usepackage{adjustbox}
\usepackage[english]{babel}
\usepackage{colortbl}
\usepackage{dirtree}
\usepackage{enumerate}
\usepackage[breaklinks]{hyperref}
\usepackage[utf8]{inputenc}
\usepackage{csquotes}
\usepackage{hhline}
\usepackage{lscape}
\usepackage{longfigure}
\usepackage{longtable}
\usepackage{rotating}
\usepackage{multirow}
\usepackage{multicol}
\usepackage[numbers]{natbib}
\usepackage{xspace}
\usepackage{xcolor}
\usepackage{tcolorbox}
\usepackage{bm}
\usepackage{footmisc}

\usepackage{todonotes}

\newcommand{\mypar}[1]{\smallskip\noindent\textbf{#1.}\xspace}

\newcommand{\system}{CANdito\xspace}
\newcommand{\attacktool}{CANtack\xspace}
\newcommand{\oldsystem}{CANnolo\xspace}

\DeclareCaptionType{mycapequ}[][List of equations]

\title{\system: Improving Payload-based Detection of Attacks on Controller Area Networks}

\author{Anonymous Authors
}

\author{
    \IEEEauthorblockN{Stefano Longari, Alessandro Nichelini, Carlo Alberto Pozzoli,  Michele Carminati, Stefano Zanero}
    \\
    \IEEEauthorblockA{\IEEEauthorrefmark{1}Dipartimento di Elettronica, Informazione e Bioingegneria, Politecnico di Milano, Milan (Italy)
    \\
    \{stefano.longari, michele.carminati, stefano.zanero\}@polimi.it
    \\
    \{carloalberto.pozzoli, alessandro.nichelini\}@mail.polimi.it 
    }
    \\

}

\begin{document}
\maketitle

    \begin{acronym}[CAN-ID]
    
    \acro{ECU}{Electronic Control Unit}
    \acro{CAN}{Controller Area Network}
    \acro{CAN-ID}{\acs{CAN} identifier}
    \acro{USB}{Universal Serial Bus}
    \acro{TPMS}{Tyre Pressure Monitoring Systems}
    \acro{IDS}{Intrusion Detetion System}
    \acro{NN}{Neural Network}
	\acro{SVM}{Support Vector Machine}
	\acro{ML}{Machine Learning}
	\acro{MSE}{Mean Squared Error}
	\acro{SGD}{Stochastic Gradient Descent}
	\acro{RNN}{Recurrent Neural Network}
	\acro{ReLU}{Rectified Linear Unit}
	\acro{ELU}{Exponential Linear Unit}
	\acro{LSTM}{Long Short-Term Memory}
	\acro{GRU}{Gate Recurrent Unit}
	\acro{DNN}{Deep Neural Network}
	\acro{DoS}{Denial of Service}
	\acro{CRC}{Cyclic Redundancy Check}
	\acro{SOF}{Start of Frame}
	\acro{EOF}{End of Frame}
	\acro{RTR}{Remote Transmission Request}
	\acro{IDE}{Identifier Extension Bit}
	\acro{DLC}{Data Length Code}
	\acro{ACK}{Acknowledgement}
	\acro{IFS}{Inteframe Space}
	\acro{IDS}{Intrusion Detection System}
	\acro{VAR}{Vector Auto Regression}
	\acro{AR}{Auto Regressive}
	\acro{DBC}{Communication Database for \acs{CAN}}
	\acro{OCSVM}{One Class \acs{SVM}}
	\acro{AUC}{Area Under Curve}
	\acro{ROC}{Receiver Operating Characteristic}
	\acro{DNN}{Deep Neural Network}
	\acro{GAN}{Generative Adversarial Network}
	\acro{FPR}{False Positive Rate}
	\acro{FNR}{False Negative Rate}
	\acro{DR}{Detection Rate}
	\acro{CPU}{Central Processing Unit}
	\acro{GPU}{Graphics Processing Unit}
	\acro{FSA}{Final State Automaton}
	\acro{API}{Application Programming Interface}
	\acro{TP}{True Positive}
	\acro{TN}{True Negative}
	\acro{FN}{False Negative}
	\acro{FP}{False Positive}
	\acro{TPR}{True Positive Ration}
	\acro{FPR}{False Positive Ratio}
	\acro{DR}{Detection Rate}
	\acro{MCC}{Matthews Correlation Coefficient}
	\acro{TTP}{Testing Time per Packet}
    \end{acronym}
    
    \acrodefplural{IDS}{Intrusion Detection Systems}
    \acrodefplural{NN}{Neural Networks}
    \acrodefplural{RNN}{Recurrent Neural Networks}
    \acrodefplural{ECU}{Electronic Control Units}
    \acrodefplural{CRC}{Cyclic Redundancy Checks}
    \acrodefplural{CAN-ID}{\acs{CAN} identifiers}

\begin{abstract}

Over the years, the increasingly complex and interconnected vehicles raised the need for effective and efficient Intrusion Detection Systems against on-board networks. In light of the stringent domain requirements and the heterogeneity of information transmitted on Controller Area Network, multiple approaches have been proposed, which work at different abstraction levels and granularities. Among these, RNN-based solutions received the attention of the research community for their performances and promising results. 

In this paper, we improve CANnolo, an RNN-based state-of-the-art IDS for CAN, by proposing CANdito, an unsupervised IDS that exploits Long Short-Term Memory autoencoders to detect anomalies through a signal reconstruction process. We evaluate CANdito by measuring its effectiveness against a comprehensive set of synthetic attacks injected in a real-world CAN dataset. We demonstrate the improvement of CANdito with respect to CANnolo on a real-world dataset injected with a comprehensive set of attacks, both in terms of detection and temporal performances.
\end{abstract}

\section{Introduction}
\label{sec:intro}

In the last decades, vehicles have become more complex, especially for what concerns their electronics~\cite{longari2019secure}. Car manufacturers nowadays implement entertainment and autonomous drive-related technologies. As a result, the number of \acp{ECU} grew to reach more than one hundred units in the most complex vehicles. 
This evergrowing complexity, however, raises security risks, as firstly demonstrated by Koscher and Checkoway in~\cite{checkoway2011comprehensive, koscher2010experimental}, allowing the attacker to gain control of the vehicle's functionalities, even remotely.
To manage such risks, the scientific research community is focusing its effort on developing countermeasures and security solutions, amongst which intrusion detection techniques for \ac{CAN} have proven effective. \acfp{IDS} for vehicular systems analyze the stream of packets and monitor the events on on-board networks for signs of intrusions. Among these, machine learning-based, and in particular RNN-based solutions, have proven to be effective in recognizing anomalous behavior~\cite{taylor2017anomaly,longari2020cannolo}. 

Based on the approach and the results of \oldsystem~\cite{longari2020cannolo}, in this paper, we propose \system, an RNN-based, unsupervised \ac{IDS} that exploits \ac{LSTM} autoencoders to detect anomalies through a signal reconstruction process. After a preprocessing stage, it learns the legitimate signal behavior through an LSTM-based autoencoder. Then, it computes the anomaly score for each CAN ID based on their reconstruction error. In particular, we improve the overall architecture and lighten \oldsystem computational requirements to meet real-world timing constraints of the automotive domain. 

We prove the effectiveness of \system by conducting experiments on a real dataset of CAN traffic augmented with a set of synthetic but realistic attacks.
With respect to existing works, we consider a broader spectrum of attacks and implement a tool to inject them into real-world CAN traffic logs. This tool is available at~\url{url.to.be.released.once.published}. We demonstrate that \system outperforms its predecessor \oldsystem, with improved detection rates and a reduction of more than 50\% of the timing overheads. 

In summary, our contributions are the following:
\begin{itemize}
    \item We improve \oldsystem with \system, an RNN-based, unsupervised \ac{IDS} that exploits \ac{LSTM} autoencoders to detect anomalies through a signal reconstruction process.
    \item \attacktool, a tool to generate and inject synthetic attacks in real datasets, which can be used as a benchmarking suite for IDS in the automotive domain.  
    \item An evaluation of \system from the point of view of detection and timing performances on a more comprehensive dataset with respect to state-of-the-art works. 
\end{itemize}

The remainder of the paper is structured as follows: 

In Section~\ref{sec:background} we provide an overview of the security issues of the CAN protocol relevant for our dissertation, while in Section~\ref{sec:related} we present a discussion on related works. These two sections together constitute the motivation for our work. In Section~\ref{sec:candito} we present \system, discussing our approach and its improvements in relation to \oldsystem. In Section~\ref{sec:attacktool} we present \attacktool, the tool that we use to generate the attacks on which we evaluate our system against \oldsystem. The experimental validation, including the setup and results of the systems, is presented in Section~\ref{sec:experiments}. Finally, in Section~\ref{sec:conclusion} we sum up our contributions and provide insights on promising research directions.
\section{Primer on CAN Security}
\label{sec:background}

\ac{CAN} has been the de-facto standard communication protocol for vehicular on-board networks since the '80. In brief, it is a bus-based multi-master communication protocol whose data packets are composed mainly of an ID and a payload (and a set of flags and control fields). The ID field (11 or 29 bits long) defines the meaning of the payload of the packet and is the first field of the packet. Each node may send multiple IDs, but an ID can be sent by only one node. Arbitration is implemented at a physical level: each node that wants to send a packet at a specific time starts writing the ID simultaneously on the bus. Given the physical properties of \ac{CAN}, a lower ID overwrites a higher one, and the overwritten node assumes to have lost arbitration. The payload is up to 8 bytes long for \ac{CAN} and up to 64 bytes long for CAN-FD. The meaning and structure of the payload of a packet with a specific ID are fixed in a specific on-board network (e.g., we may define that in our on-board network ID 0x28's payload first two bytes send the sensor data regarding the speed of the vehicle, and the second two bytes the sensor data of the RPM. We expect this structure in every packet on the bus with ID 0x28). For further details on the \ac{CAN} specification, we refer the reader to~\cite{boschcanv2}.

\mypar{Attacks on CAN}
\ac{CAN} security weaknesses are nowadays well known and discussed in multiple works~\cite{young2019survey,buttigieg2017security}. As demonstrated by Miller and Valasek in~\cite{miller2013adventures,miller2015remote}, one of the most common known vulnerabilities derives from the lack of authentication of messages on \ac{CAN}. A node should not be allowed to send IDs that it does not own, but there is no mechanism to enforce this rule. Therefore, an attacker that takes control of an \ac{ECU} that has access to a \ac{CAN} bus can ideally send any ID and payload. In worst-case scenarios, the attacker is also capable of silencing the owner of the packet to avoid conflicts, as presented in~\cite{longari2019copycan}. 
Given the \ac{IDS} nature of the work at hand, we present the capabilities of the attacker through the effects of its actions on the payload and flow of packets on the bus: A weak attacker may \textbf{inject} forged packets with one or multiple specific IDs on the bus, without silencing their owner. In this situation, the receivers may or may not consider the attacker's packets valid due to the incongruities on the bus. To solve this conflict, a stronger attacker may silence the owner and then forge packets with its ID, leading to a \textbf{masquerade} attack. Such masquerade attack can be implemented with or without consideration of the existence of an \ac{IDS} checking the bus. If it is considered, the attacker may want to implement a \textbf{replay} attack, where she/he does not create a new payload but repeats a payload previously captured on the bus, or a \textbf{seamless change} attack, where the attacker drives a signal from its current value to a tampered one by changing it slowly through multiple packets. If it is not considered, an attacker may want to study the effects of various payloads and IDs on the system, implementing a \textbf{fuzzing attack}. Finally, an attacker may have the only goal to silence a node without replacing it, creating a \textbf{drop} attack. As further discussed in Section~\ref{sec:attacktool} when we present our attack tool, we generate datasets that consider these attacks and evaluate the systems against all of them.
\section{Related Works}
\label{sec:related}

Intrusion detection for automotive ecosystems has been proposed both off-board, for VANETs~\cite{yang2019tree,garg2020multi}, both for on-board and intra-vehicle communication. Intra-vehicle intrusion detection is a wide research topic for which we refer the reader to Al-Jarrah et al.~\cite{al2019intrusion} that present a comprehensive survey of intra-vehicle \acp{IDS}, which can be divided into flow-based, payload-based, and hybrid. Flow-based \acp{IDS} (e.g., ~\cite{SongFrequency2016, taylor2015frequency, seo2018gids}) monitor the CAN bus, extract distinct features as message frequency or packet inter-arrival time, and use them to detect anomalous events without inspecting the payloads of the messages. On the contrary, payload-based \acp{IDS} (e.g.,~\cite{longari2020cannolo,kang2016intrusion,hanselmann2020canet}) examine the payload of \ac{CAN} packets (usually only data frames). Finally, hybrid \ac{IDS} (e.g., ~\cite{Zhang2018ATD, marchetti2016evaluation}) are a combination of the previous two techniques. In this section, we focus on payload-based \acp{IDS} and their characteristics since they are more related to \system. 

Kang and Kang~\cite{kang2016intrusion} propose a supervised payload-based \ac{IDS} based on \ac{DNN}.  The input feature does not use the entire payload, but only the \textit{mode information}, which represents the command state of an ECU, and the \textit{value information}, which represents the value of the mode (e.g., wheel angle or speed). 
The approach is tested on a synthetic dataset created with OCTANE~\cite{borazjani2014octane} and shows a \ac{FPR} of 1.6\% and a \ac{FNR} of 2.8\%. Nevertheless, this work has some critical limitations: the \ac{FPR} is high, the testing time between 2ms and 5ms (on a not specified hardware) makes it suitable for real-time detection only on a limited set of CAN IDs (e.g., in our dataset the average packet interarrival time on all CAN IDs was 0.4ms) and not on the entire network traffic, it is a supervised method, able to detect the specific type of attacks it is trained on, but it cannot generalize to detect novel anomalous events.

Longari et al.~\cite{longari2020cannolo} propose an unsupervised \ac{IDS} based on a \ac{LSTM} autoencoder, and represents the starting point of \system. This approach shows good performances on various data field anomalies, with an overall \ac{AUC} of 0.9677, higher than state-of-the-art solutions. However, it has some significant limitations: slow computation time, and it has been tested only on a limited set of CAN IDs. In this work, we aimed at improving \system and studying a broader set of CAN IDs. 

Hanselmann et al. proposed CANet~\cite{hanselmann2020canet}, an \ac{IDS} based on an autoencoder that has a separate input \ac{LSTM} network for the traffic of each CAN ID.
The inputs of each \ac{LSTM} sub-network are the signals of the correspondent CAN IDs rescaled with a signal-wise 0-1 normalization. This approach has the unique advantage compared to other state-of-the-art methods of potentially detecting a more significant number of anomalies thanks to the fact that it considers every CAN ID at the same time. Nonetheless, the time to test each packet has proven to be too long in order to use this method for real-time detection on a large set of CAN IDs or in series with other \acp{IDS}. Our implementation on Tensorflow 2.1 and tested on a system with an Intel Core i5-8300H \ac{CPU} and an NVIDIA GeForce GTX 1050 \ac{GPU} showed a testing time per packet of 2.5ms, which is much higher than the average time between packets of 0.4ms of our dataset.

\subsection{Motivation} 

The main limitation of current state-of-the-art approaches is that, while different methods work well on different problems, none of them can achieve results that are good enough on any anomaly and, at the same time, provide fast enough results to process the network's traffic in real-time. The way this limitation compels a system largely depends on the different types of \ac{IDS} approach adopted. Generally, flow-based approaches can provide fast predictions while being limited to the detection of specific kinds of vulnerabilities, while payload-based approaches have a broader scope, but it is often a problem to make them work in real-time on the total traffic. Moreover, the traffic on different CAN IDs has different characteristics, but the current state-of-the-art methods rarely consider them in order to provide better results.

\section{\system}
\label{sec:candito}

In this section, we describe \system, an improved version of \oldsystem~\cite{longari2020cannolo}, a state-of-the-art IDS for CAN that exploits \ac{RNN}-based autoencoders.  
\ac{RNN}-based architectures are effective in modeling time-series and have been successfully proposed for \ac{CAN} traffic anomaly detection~\cite{taylor2017anomaly}. 
On the other hand, autoencoders do not require a labeled training dataset since target signals are generated automatically from the input sequence. Moreover, being unsupervised, they learn a model of the legitimate network traffic and not specific anomalies, making them able to potentially recognize novel attacks.  

\begin{figure}
	\centering
	\includegraphics[width=\linewidth]{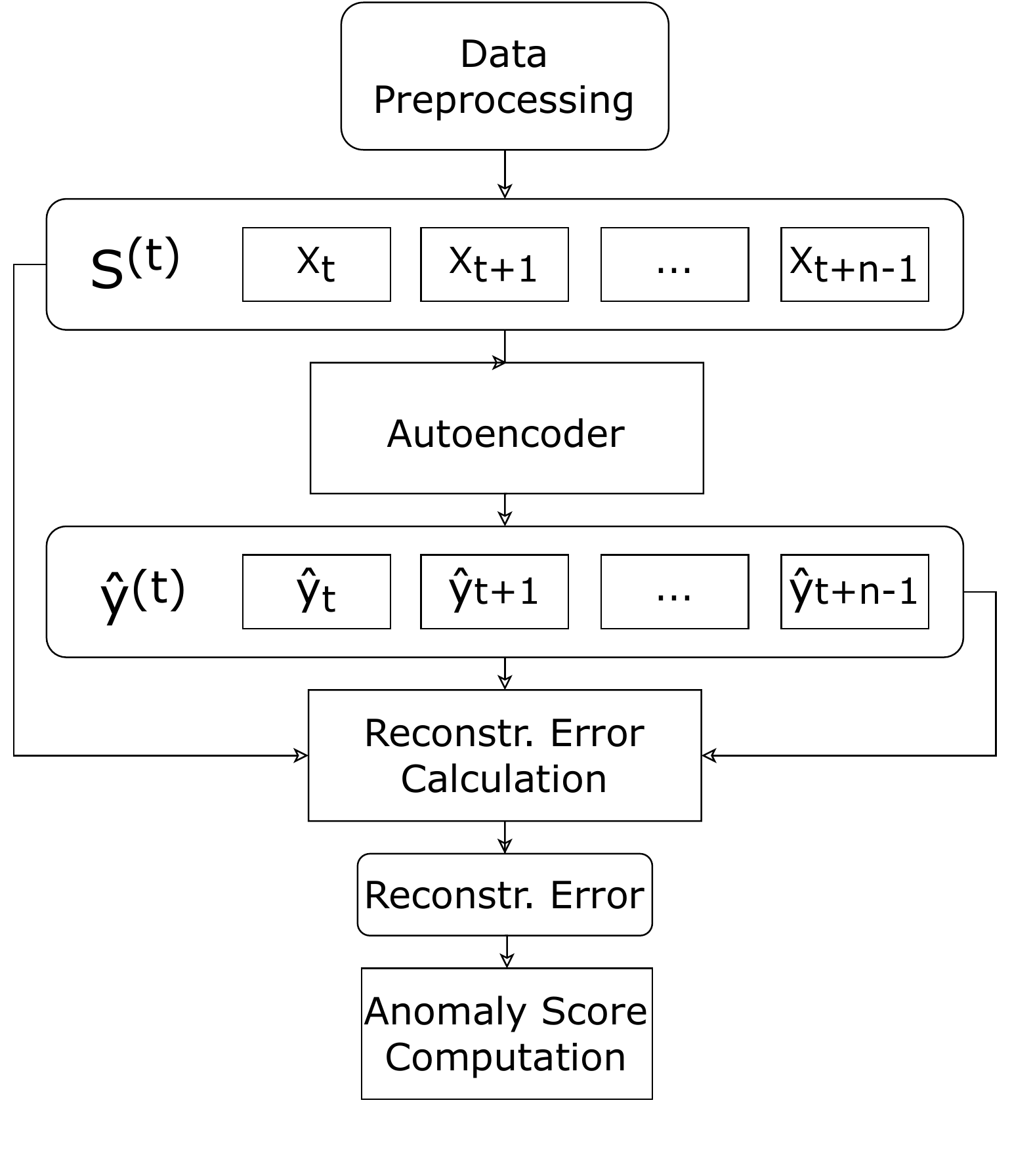}
	\caption{Overview of \system's detection process. }
	\label{fig:autoencoderoverview}
\end{figure}

Figure~\ref{fig:autoencoderoverview} shows an overview of the architecture of the system, which works by reconstructing time series of CAN packets for each ID and computes their anomaly score based on the reconstruction error. The effectiveness of reconstructing the signal (opposed to predicting the successive one) has been demonstrated by Malhotra et al.~\cite{malhotra2016lstm}, which uses LSTM encoder-decoder architectures as reconstructors to detect anomalies in multi-sensor contexts.
It comprises three modules: a data preprocessing module, an LSTM-based autoencoder, which learns the legitimate signal behavior, and an anomaly detector, which compares the reconstruction errors.  

\subsection{Data Preprocessing}

\begin{table}
\centering
\begin{tabular}{@{}ll@{}}
0x1D0 & 0x201 \\ \midrule
\multicolumn{1}{|l|}{Wh.Speed FL (phys,0-15)} &
  \multicolumn{1}{l|}{Gas (phys, 0-15)} \\ 
\multicolumn{1}{|l|}{Wh.Speed FR (phys,16-31)} &
  \multicolumn{1}{l|}{Gas2 (phys, 16-31)} \\ 
 \multicolumn{1}{|l|}{Wh.Speed RL (phys,32-47)} &
  \multicolumn{1}{l|}{\textit{unlabelled} (n/a, 32-33)} \\ 
 \multicolumn{1}{|l|}{Wh.Speed RR (phys,48-63)} &
  \multicolumn{1}{l|}{Counter (counter, 34-35)} \\ 
 \multicolumn{1}{|l|}{} &
  \multicolumn{1}{l|}{Checksum (CRC,36-39)} \\ \bottomrule
\end{tabular}%
\caption{Examples of READ-parsed CAN IDs, from Marchetti and Stabili's~\cite{marchetti2018read}}
\label{tab:readexample}
\end{table}

The first module of \system builds the input sequences by applying the READ algorithm~\cite{marchetti2018read} and associating to each CAN ID the corresponding ranges of the signals. Using this information, the payloads of each packet are converted into the list of their signals rescaled in the [0-1] range, excluding constant bits, counters, and \acp{CRC}. In Table~\ref{tab:readexample} an example of how two IDs and their payloads may be cataloged by READ.
Our resulting input sequence is composed by 
a matrix \textit{n} x \textit{k}, where \(\mathit{n}=40\) is the dimension of the window of CAN packets, \textit{k} is the number of signals per packet (rescaled in the [0-1] range).

\subsection{LSTM-based autoencoder} 

\begin{figure}
    \centering
    \includegraphics[width=\columnwidth]{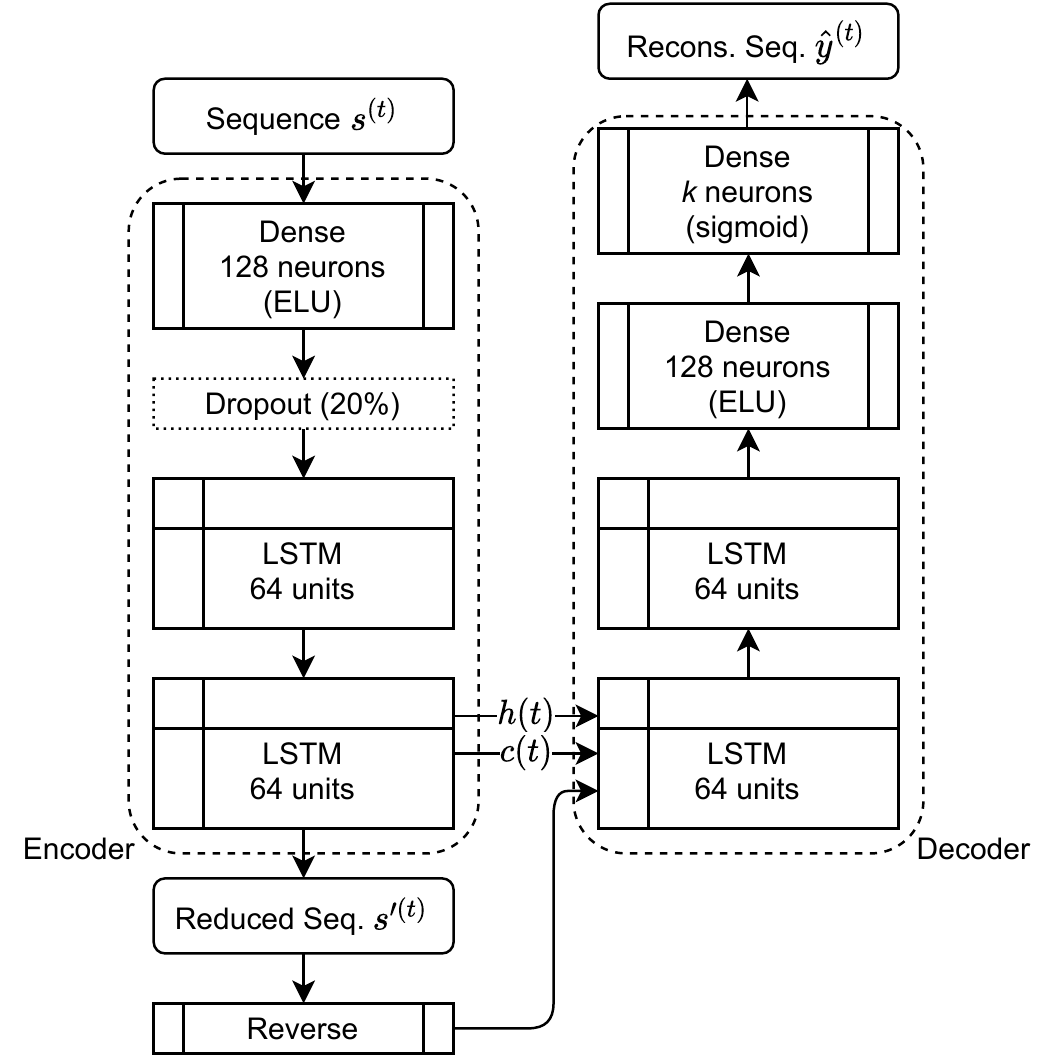}
    \caption{The architecture of \system's autoencoder}
    \label{fig:autoencoderids}
\end{figure}

As shown in Figure~\ref{fig:autoencoderids}, the second module of \system is based on an autoencoder whose encoder and decoder layers are implemented with two recurrent \acp{LSTM}.

The encoder is  composed of a dense layer, which consists of $128$ units with an \ac{ELU}~\cite{clevert2015fast} activation function. The dense layer is followed by a $20\%$ dropout layer and two recurrent LSTM layers with $L=64$ units each. The cell and hidden states of the last \ac{LSTM} layer of the encoder are used to initialize the states of the first \ac{LSTM} layer of the decoder. The output of the encoder is reversed before being fed into the decoder. Symmetrically, the decoder consists of two recurrent LSTM layers with $64$ units each, a dense layer, consisting of $128$ units with \ac{ELU} activation function, 
and a second output dense layer with \textit{k} units with sigmoid activation function (to scale the results to the [0,1] interval). 

\subsubsection{Training and Tuning}

The input data of the training and threshold calculation processes are composed by sliding one time step at a time a window of $\mathit{n}$ packets, while for the testing process, in order to have a lightweight detection process, the windows of packets do not overlap.
A dataset consisting only of legitimate data sequences has been used as the baseline to establish the `normal' behavior for any given CAN ID. In particular, we perform training and validation by reconstructing legitimate traffic data and minimizing the reconstruction error between a given source sequence $s^{(t)}$ and a target sequence $y^{(t)}$. Also, we make use of an untampered dataset to perform early stopping and a dataset injected with our attack tool (presented in Section~\ref{sec:attacktool}) to evaluate the performance of the model and, consequently, tune hyperparameters. 
The loss function of choice is \ac{MSE}. The optimizer of choice is Adam~\cite{kingma2014adam} with a learning rate of $0.001$. The model of each CAN ID has been trained for a maximum of 50 epochs with an early stopping with patience 5 (i.e., training is stopped before the maximum number of epochs if the validation accuracy does not improve for 5 consecutive epochs).

\subsection{Anomaly score computation}

The third module of \system works in an unsupervised fashion by computing the distance between the reconstruction error and the expected normal distribution computed during training. The anomaly score indicates the likelihood of the test sequence to be anomalous. 

Each window of $\mathit{n}$ packets is fed into the trained autoencoder and an anomaly score is assigned to each window as the squared l2-norm of the reconstruction error:
\begin{equation} \label{squarederror}
    e^{(t)}=\|\hat{y}^{(t)}-s^{(t)}\|^2_2
\end{equation}
The chosen detection threshold is, as proposed by Hanselmann et al.~\cite{hanselmann2020canet}, the 99.99 percentile of the scores. The score for each new testing window is calculated with Equation~\ref{squarederror}. The windows whose score is greater than the threshold are marked as anomalies.

\subsection{Improvements from \oldsystem}

As previously stated, \system is based on \oldsystem~\cite{longari2020cannolo}, in light of its promising results in the detection of attacks on CAN. 
In particular, \system required an in-depth study of \oldsystem. To do so, we implemented \oldsystem from scratch and tested it on our extended dataset. This evaluation brought the improvements described below.      

\mypar{Generalization and computation improvements} The number of artificial neurons of all layers has been halved. This reduction in dimensions has a twofold scope. The networks of several CAN IDs are prone to overfitting, reducing the dimension of the layers mitigated this problem. Moreover, reducing the dimension of the layers has a substantial impact on the improvement of the computation times of the network.

\mypar{Overfitting and vanishing gradient mitigation} The activation functions of the two symmetric dense layers of the encoder and decoder have been modified from hyperbolic tangent to \ac{ELU}. This was proven to mitigate the vanishing gradient problem and to improve the generalization capabilities of the network~\cite{clevert2015fast,hanselmann2020canet}.

\mypar{Input bloat reduction} The inputs of \oldsystem are bit-strings composed of the condensed notation of the packets (i.e., excluding constant bits). In our solution, we also exclude signals marked by READ as counters or \acp{CRC}. In fact, counters and \acp{CRC} do not carry relevant information for the reconstruction module, as demonstrated by the fact that considering them did not improve the effectiveness of the system. Moreover, \system's input is not composed of bit-strings but by each signal detected by READ rescaled in the [0-1] range. This significantly reduces the input dimensions, further lowering computing times, while comparative tests with the two input methods did not show meaningful effects on detection performances. 

\mypar{Underfitting mitigation} The output sequence of the encoder is reversed before being fed into the decoder. This operation is meant to help the network reconstruct the target sequence, which is also reversed as suggested by Sutskever et al.~\cite{sutskever2014sequence}. While for some CAN IDs the network results are similar with or without reversing the encoder output, other CAN IDs networks are affected by severe underfitting if the encoder output is not reversed. The same CAN IDs networks perform well after the change. 
  
\mypar{Lower computational requirements} To lower the computational effort, we do not feed the reconstructed sequence back into the decoder. Evaluations at design time did not show improvements in detection performances between the two implementations.

\mypar{Anomaly score computation improvement} \oldsystem uses the Mahalanobis distance between the reconstruction error of the window under evaluation and the distribution of errors in untampered scenarios. While such distance has been considered for the anomaly scores computation, it has underperformed with our model. Instead, we opted to compute the anomaly score as the squared l2-norm of the reconstruction error, with a $99.99$ percentile of the scores as a detection threshold, as suggested in~\cite{hanselmann2020canet}.

\section{\attacktool}
\label{sec:attacktool}

We designed \attacktool to have an easy, partially automated way to consistently generate different types of anomalies in our datasets while starting from datasets of real \ac{CAN} traffic. The tool is available at \footnote{\label{cantackurl}\attacktool url: url.to.be.released} alongside instructions on how to use it. The output of the tool is a dataset structured similarly to the ReCAN dataset~\cite{zago2020recan}, but with an additional \textit{isTampered} column, which indicates whether a log entry has been tampered with or not. 

The tool allows choosing between the different attack implementations, which enable to deploy all the attacks presented in Section~\ref{sec:background}. Moreover, all different types of attacks (except for cases in which this does not make sense, i.e., \textit{drop} and \textit{DoS}) can be either performed in an \textit{injection} fashion (i.e., without modifying the original packets of the traffic and specifying an injection rate) or in a \textit{masquerade} fashion (i.e., substituting the original packets with the tampered ones).
We proceed to present the list of attacks and their user-defined parameters. Note that for all attacks, the user needs to set the ID to tamper and the beginning time for the attack in seconds.

\mypar{Basic injection} The user can specify a payload and a number of tampered packets. The tool injects or replaces a number of packets with the new tampered payload.

\mypar{Progressive injection} As above, the value of every single payload can be specified.

\mypar{DoS} The network is flooded with packets with ID ''$0$'' for the specified duration. It is possible to define the percentage of the bus to fill. 

\mypar{Drop} A set amount of messages from the given ID are deleted from the dataset.

\mypar{Fuzzy} The payloads are injected or tampered with random values. It is also possible to choose a bit range in which to fuzzy a value, e.g., simulating the fuzzing only of a sensor data.

\mypar{Replay} The payloads are values sniffed from the dataset. To implement this attack, it is necessary to define the initial sniffing time and whether to partially randomize the position of the first copied packet among the sniffed ones. Moreover, a series of replacements can be specified to modify only some bit-ranges with other values. These values can be set with the following options: 
\begin{itemize}
\item \textbf{payloads} replaces the bit-ranges with explicitly defined data;
\item \textbf{fuzzy} randomizes the bit-range;
\item \textbf{min} and \textbf{max} respectively find the minimum and maximum value (considered as an integer) registered in the dataset for that bit-range;
\item \textbf{seamless change} defines a final value to reach and increases or decreases the bit-range from the value read in the last untampered line to the chosen one;
\item \textbf{counter} increases the values of the bit-range by one per packet, in a counter-like fashion. 
\end{itemize}

\section{Experimental validation}
\label{sec:experiments}

In this experiment, we compare \system with \oldsystem~\cite{longari2020cannolo} from both the point of view of the detection and the temporal performances. 
In particular,
the experiment consists of evaluating the performance of \oldsystem and \system over different datasets tampered with our novel attack tool~\ref{sec:attacktool}. 
We evaluate the systems both on the entire dataset (see Table~\ref{tab:resultslong}) and on the subset of CAN IDs on which \oldsystem is considered optimal by its authors (see Table~\ref{tab:resultsshort}). 
\oldsystem's implementation has been kept almost untouched, threshold computation criterion included. This consideration must be explicitly made since the authors used as a comparison metric for their experiment just the \ac{AUC} without focusing over the threshold computation criterion, which was demonstrated to be sub-optimal. 

We evaluate the detection performances of the systems under analysis by considering the most common metrics used to evaluate unbalanced datasets. Specifically, we make use of \ac{DR}, \ac{FPR}, F1 score, \ac{MCC}. Moreover, to evaluate the timing performances, we use the \ac{TTP}.

In order to comprehend the results of the evaluated \acp{IDS}, one clarification is needed: 
We aim to detect anomalous behavior. This defines the scope of our \ac{IDS}, which is not to find and suppress one single anomalous message but to detect the presence of anomalies or intrusions and raise an alarm. Although detecting all and only attack packets would be optimal, this does not carry any evident advantage for the IDS in the vast majority of cases. The only exception occurs when the final goal of the IDS is that of deleting the attack messages, which is a drastic measure generally not considered given the safety issues of vehicular networks.
Therefore, we do not evaluate the results per packet, but by using a window of dimension $n=40$ (the same dimension of the input window of \system) and consider a window non-anomalous only if none of its 40 packets is anomalous. It is enough to find one positive prediction to consider the whole window anomalous.

Both the systems have been tested by serving the dataset split in windows of pre-defined size as it should happen at runtime, instead of testing the entire sequence in one batch; doing so permits to have a measure of the testing time that is more accurate.

\subsection{Datasets}
\label{ssec:dataset}

To build and evaluate \system, we chose the ReCAN dataset~\cite{zago2020recan}, a public dataset of CAN logs retrieved in real-world scenarios. We then select the C-1 sub-dataset, which has been retrieved from multiple test drives of an Alfa Giulia. 
In more detail, we use sub-datasets 1, 2, 6, 8, and 9 to train the model, sub-dataset 4 to calculate thresholds, sub-dataset 5 for validation and hyper-parameter tuning, and finally, sub-dataset 7 for testing.

We then use our attack tool to generate the attack datasets. All the datasets are generated starting from sub-dataset 7 of the ReCAN C-1 dataset. The datasets have the goal of building attacks as presented in our background Section~\ref{sec:background}. All the datasets, their details, and implementation parameters are available on the \attacktool webpage\footref{cantackurl}.

\mypar{Injection Dataset} This dataset contains generic injection attacks, which consists of added packets on the network, leaving all packets already present in the dataset unchanged. The new packets are sniffed from previous traffic and only one physical signal is modified (recognized through the READ algorithm~\cite{marchetti2018read}), although the value changes inside the range of already existing values of the signal. This is done in an attempt to comply as much as possible with the behavior of the ID and increase the difficulty of detection. The packets are added at 20 times the frequency of the original packet and continue for a sequence of 50 packets.

\mypar{Drop Dataset} This dataset simulates the event where an attacker turns off an \ac{ECU} or its \ac{CAN} controller. The attack consists of removing a sequence of valid packets from the original traffic. Twenty-five consecutive packets are removed each time. 

\mypar{Masquerade Dataset} This dataset contains generic masquerade attacks that would not be detectable only through frequency-based or rule-based features. The modified packets are sniffed from previous traffic and one or more physical signals are modified in the same fashion as the injection dataset. Moreover, the anomalous packets maintain the same timestamps as the packet they replace, alongside its ID. Each anomalous sequence has a length of 25 packets.

\mypar{Fuzzed Dataset} This dataset represents the event where an attacker is testing random values of signals in order (usually) to trigger unexpected behavior. The attack is made in a masquerade fashion (the original packets are removed and replaced), but only the bits included in some of the signals are fuzzed, while, for example, constant bits are left untampered. As above, 25 consecutive packets are removed each time. As above, each anomalous sequence has a length of 25 packets.

\mypar{Seamless Change Dataset} This dataset contains masquerade attacks that attempt to evade detection by changing the payload of the packets progressively until the desired value is reached. The physical values in the tampered ID have to be at least 4 bits long. The new packets are sniffed from previous traffic and only one physical signal is modified. As above, each anomalous sequence has a length of 25 packets.

\mypar{Full Replay Dataset} This dataset contains masquerade attacks that attempt to evade detection by copying exact sequences on the bus. No additional check is made while generating the attacks. Consequently, there is no warranty that the new sequence is taken from a moment where the condition of the car is very different, lowering the detection capabilities but also the actual effects of the attacks. This dataset is primarily interesting to compare the ability of \ac{IDS}to detect anomalous sequences that are perfectly valid in a different context. As above, each anomalous sequence has a length of 25 packets.

\subsection{Results}
\label{ssec:results}

\begin{table}
\Large
\resizebox{\columnwidth}{!}{
\begin{tabular}{lllllll}
\toprule
 Dataset &  & DR & FPR & F1 & MCC & TTP \\ 
\midrule

\multirow{2}{*}{Masq.} &

\system &
\textbf{.9258} &
.0081 &
\textbf{.9505} &
\textbf{.9336} &
\textbf{.4560ms} \\ 

& \oldsystem &
.6477 &
\textbf{.0029} &
.7823 &
.7502 &
1.0821ms \\ 
  
\midrule
  
\multirow{2}{*}{Fuzzy} &

{\system} &
  \textbf{.9989} &
  .0081 &
  \textbf{.9886} &
  \textbf{.9844} &
  \textbf{.4513ms}  \\
  
&   \oldsystem &
  .9541 &
  \textbf{.0029} &
  .9724 &
  .9629 &
  1.0628ms 
  \\ \midrule
 
\multirow{2}{*}{Seam.} &

 {\system} &
  \textbf{.8972} &
  .0079 &
  \textbf{.9345} &
  \textbf{.9143} &
  \textbf{.4613ms} \\ 
  
&  {\oldsystem} &
  .7481 &
  \textbf{.0029} &
  .8518 &
  .8224 &
  1.0701ms \\ \midrule
  
\multirow{2}{*}{Replay} &

 {\system} &
  \textbf{.5820} &
  .0080 &
  .7254 &
  .6909 &
  \textbf{.4483ms} \\ 
  
&  {\oldsystem} &
  .5801 &
  \textbf{.0029} &
  \textbf{.7304} &
  \textbf{.7024} &
  1.0373ms \\ 
  \bottomrule
\end{tabular}
}
\caption{\oldsystem vs \system performances, tested over the masquerade, fuzzy, seamless change, and full replay datasets. Only CAN IDs recommended for testing by \oldsystem's authors have been taken into consideration. }
\label{tab:resultsshort}
\end{table}

\begin{table}
\Large
\resizebox{\columnwidth}{!}{%
\begin{tabular}{llllllll}
\toprule
 Dataset &  & DR & FPR & F1 & MCC & TTP \\ \midrule
 
\multirow{2}{*}{Masq.} &
{\system} &
  \textbf{.9066} &
  .0265 &
  \textbf{.9149} &
  \textbf{.8854} &
  \textbf{.4560ms} \\
& {\oldsystem} &
  .6446 &
  \textbf{.01451} &
  .7648 &
  .7219 &
  1.0821ms \\ \midrule

\multirow{2}{*}{Fuzzy} &
{\system} &
  \textbf{.9954} &
  .0231 &
  \textbf{.9683} &
  \textbf{.9567} &
  \textbf{.4513ms} \\ 
& {\oldsystem} &
  .9243 &
  \textbf{.0145} &
  .9418 &
  .9207 &
  1.0628ms \\ \midrule
  
\multirow{2}{*}{Seam.} &
{\system} &
  \textbf{.8850} &
  .0263 &
  \textbf{.8996} &
  \textbf{.8684} &
  \textbf{.4613ms} \\ 
& {\oldsystem} &
  .7320 &
  \textbf{.0149} &
  .8238 &
  .7865 &
  1.0701ms \\ \midrule

\multirow{2}{*}{Replay} &
{\system} &
  \textbf{.5723} &
  .0286 &
  .6896 &
  .6347 &
  \textbf{.4483ms} \\ 
& {\oldsystem} &
  .5534 &
  \textbf{0.0146} &
  \textbf{.6926} &
  \textbf{.6547} &
  1.0373ms \\ \midrule
  
\multirow{2}{*}{Inj.} &
{\system} &
  \textbf{.6901} &
  .0277 &
  \textbf{.3826} &
  \textbf{.4150} &
  \textbf{.4715ms} \\ 
& {\oldsystem} &
  .5490 &
  \textbf{.0194} &
  .3800 &
  .3878 &
  1.0410ms \\ \midrule
  
\multirow{2}{*}{Drop} &
{\system} &
  .2300 &
  .0238 &
  .3474 &
  .3346 &
  \textbf{.4714ms} \\ 
& {\oldsystem} &
  \textbf{.5085} &
  \textbf{.0191} &
  \textbf{.6420} &
  \textbf{.6089} &
  1.1428ms \\ \bottomrule
\end{tabular}%
}
\caption{\oldsystem vs \system performances, tested over the masquerade, fuzzy, seamless change, and full replay datasets. All CAN IDs have been taken into consideration. }
\label{tab:resultslong}
\end{table}

As shown in Table~\ref{tab:resultsshort} and Table~\ref{tab:resultslong}, our solution is not only twice as fast as \oldsystem in providing the predictions, but it is also more effective in almost all the considered attack scenarios. In particular, in Table~\ref{tab:resultslong} we show the results on the entire dataset, while in Table~\ref{tab:resultsshort}, we show the results of the two models on the same set of CAN IDs (i.e., $0DE$, $0EE$, $0FB$, $0FC$, $0FE$, $0FF$, $1F7$, $1FB$, $11C$, $100$, $104$, $116$). 


Focusing on payload-based anomalies, \system generally outperforms \oldsystem on both the entire dataset and on the subset of the dataset composed of the selected CAN IDs.
For the Masquerade dataset, \system performs evidently better, with an F1 score and MCC both over .15 point higher than \oldsystem. We explain the better performances obtained with the different approaches used to compute the detection threshold. For the Fuzzy datasets, \system shows similar performances to \oldsystem, with a higher \ac{DR}, F1-score, and \ac{MCC}. 
For the Seamless Change dataset, the better performances of our new architecture are more evident. In fact, both the F1 score and \ac{MCC} improvements range between .07 and .09. For the Full replayed dataset, the original model is slightly more effective, but the performances of the two systems are comparable. However, in light of the stringent requirements of the automotive domain, where the lack of computational power is critical~\cite{maffiola2021goliath}, \system is preferable to \oldsystem since it provides detection results in less than half of the time. 

As expected, both systems perform poorly on flow-based anomalies (i.e., Injection and Drop datasets) since they implement payload-based detection and do not detect changes in the frequency of packets arrivals. 

It is interesting to note that CAN ID $0x1E340000$ is responsible for the 18\% of overall false positives and features a very different behavior between the train set and the test set with the presence of flipping bits that were static for the entire duration of the training set. 
This said, it is encouraging to know that a large part of the \ac{FPR} depends on a small set of CAN IDs because this demonstrates that future works may improve the results of finding a different classification of these ''pathological'' CAN IDs. Another possible alternative that is not particularly time-consuming, considering the small dimension of this set of CAN IDs is to perform a human supervised fine-tuning of the model for these specific CAN IDs on top of the automatic classification.

\subsection{Discussion}
Comparing the results of \system with the ones provided by other state-of-the-art techniques might be complicated considering that usually the results of the systems are not provided on a large set of CAN IDs; on the other hand, it is still possible to make some considerations about this topic. As we discuss in Section~\ref{sec:related}, flow-based \acp{IDS} are not able to detect any payload-based anomaly, so it is not easy to compare them with a more general approach. 

Taking into consideration payload-based approaches, besides \oldsystem~\cite{longari2020cannolo}, one of the most interesting state-of-the-art techniques is the autoencoder CANet proposed by Henselmann et al.~\cite{hanselmann2020canet}. The results provided by the authors are promising, but they are performed only on a set of 13 CAN IDs that have a high correlation between them, and this makes it difficult to compare the two systems, given that in our tests, it is easy to find both CAN IDs on which our system performs well and others on which the performances are lower. This said, the only attack on which CANet performs better is what the authors call ''playback attack'', which is similar to the attacks performed in our full replay dataset, thanks to the fact that their system is able to take into consideration the correlation between the different CAN IDs.

\section{Conclusions}
\label{sec:conclusion}

In this paper, we presented \system, an improved RNN-based and unsupervised \ac{IDS} that exploits \ac{LSTM} autoencoders to detect anomalies through a signal reconstruction process 
in \ac{CAN} traffic. We evaluated \system from the point of view of the detection and timing performances on a more comprehensive real-world dataset augmented with synthetic attacks generated with \attacktool,  a tool to generate and inject synthetic attacks in real datasets, which can be used as a benchmarking suite for IDS in the automotive domain.  
We demonstrated that \system is able to out-perform the state of the art, improving the detection rate by 11.85\% on average and reducing the timing overheads by 50\%. 
We plan to overcome \system limitation in detecting attacks that work in the frequency domain by complementing the improved detection power of the payload-based detection system presented in this work with the power of frequency-based approaches to building an end-to-end hybrid \ac{IDS} able to fully exploit all CAN IDs.


%
\IEEEpeerreviewmaketitle



\bibliographystyle{IEEEtran}
\bibliography{bibliography}

\begin{thebibliography}{10}
\providecommand{\url}[1]{#1}
\csname url@samestyle\endcsname
\providecommand{\newblock}{\relax}
\providecommand{\bibinfo}[2]{#2}
\providecommand{\BIBentrySTDinterwordspacing}{\spaceskip=0pt\relax}
\providecommand{\BIBentryALTinterwordstretchfactor}{4}
\providecommand{\BIBentryALTinterwordspacing}{\spaceskip=\fontdimen2\font plus
\BIBentryALTinterwordstretchfactor\fontdimen3\font minus
  \fontdimen4\font\relax}
\providecommand{\BIBforeignlanguage}[2]{{%
\expandafter\ifx\csname l@#1\endcsname\relax
\typeout{** WARNING: IEEEtran.bst: No hyphenation pattern has been}%
\typeout{** loaded for the language `#1'. Using the pattern for}%
\typeout{** the default language instead.}%
\else
\language=\csname l@#1\endcsname
\fi
#2}}
\providecommand{\BIBdecl}{\relax}
\BIBdecl

\bibitem{longari2019secure}
S.~Longari, A.~Cannizzo, M.~Carminati, and S.~Zanero, ``A secure-by-design
  framework for automotive on-board network risk analysis,'' in \emph{2019 IEEE
  Vehicular Networking Conference (VNC)}, 2019, pp. 1--8.

\bibitem{checkoway2011comprehensive}
S.~Checkoway, D.~Mccoy, D.~Anderson, B.~Kantor, H.~Shacham, S.~Savage,
  K.~Koscher, A.~Czeskis, F.~Roesner, and T.~Kohno, ``Comprehensive
  experimental analyses of automotive attack surfaces,'' 08 2011.

\bibitem{koscher2010experimental}
K.~Koscher, A.~Czeskis, F.~Roesner, S.~Patel, T.~Kohno, S.~Checkoway, D.~McCoy,
  B.~Kantor, D.~Anderson, H.~Shacham \emph{et~al.}, ``Experimental security
  analysis of a modern automobile,'' in \emph{2010 IEEE Symposium on Security
  and Privacy}.\hskip 1em plus 0.5em minus 0.4em\relax IEEE, 2010, pp.
  447--462.

\bibitem{taylor2017anomaly}
A.~Taylor, ``Anomaly-based detection of malicious activity in in-vehicle
  networks,'' Ph.D. dissertation, Universit{\'e} d'Ottawa/University of Ottawa,
  2017.

\bibitem{longari2020cannolo}
S.~Longari, D.~H.~N. Valcarcel, M.~Zago, M.~Carminati, and S.~Zanero,
  ``Cannolo: An anomaly detection system based on lstm autoencoders for
  controller area network,'' \emph{IEEE Transactions on Network and Service
  Management}, 2020.

\bibitem{boschcanv2}
{Robert Bosch GMBH}, ``Can specification, version 2.0,'' {Robert Bosch GmbH},
  Stuttgart, Germany, Standard, 1991.

\bibitem{young2019survey}
\BIBentryALTinterwordspacing
C.~Young, J.~Zambreno, H.~Olufowobi, and G.~Bloom, ``Survey of automotive
  controller area network intrusion detection systems,'' \emph{{IEEE} Des.
  Test}, vol.~36, no.~6, pp. 48--55, 2019. [Online]. Available:
  \url{https://doi.org/10.1109/MDAT.2019.2899062}
\BIBentrySTDinterwordspacing

\bibitem{buttigieg2017security}
\BIBentryALTinterwordspacing
R.~Buttigieg, M.~Farrugia, and C.~Meli, ``Security issues in controller area
  networks in automobiles,'' \emph{CoRR}, vol. abs/1711.05824, 2017. [Online].
  Available: \url{http://arxiv.org/abs/1711.05824}
\BIBentrySTDinterwordspacing

\bibitem{miller2013adventures}
C.~Miller and C.~Valasek, ``Adventures in automotive networks and control
  units,'' \emph{Def Con}, vol.~21, no. 260-264, pp. 15--31, 2013.

\bibitem{miller2015remote}
------, ``Remote exploitation of an unaltered passenger vehicle,'' \emph{Black
  Hat USA}, vol. 2015, no. S 91, 2015.

\bibitem{longari2019copycan}
\BIBentryALTinterwordspacing
S.~Longari, M.~Penco, M.~Carminati, and S.~Zanero, ``Copycan: An error-handling
  protocol based intrusion detection system for controller area network,'' in
  \emph{Proceedings of the {ACM} Workshop on Cyber-Physical Systems Security
  {\&} Privacy, CPS-SPC@CCS 2019, London, UK, November 11, 2019}, L.~Cavallaro,
  J.~Kinder, and T.~Holz, Eds.\hskip 1em plus 0.5em minus 0.4em\relax {ACM},
  2019, pp. 39--50. [Online]. Available:
  \url{https://doi.org/10.1145/3338499.3357362}
\BIBentrySTDinterwordspacing

\bibitem{yang2019tree}
L.~Yang, A.~Moubayed, I.~Hamieh, and A.~Shami, ``Tree-based intelligent
  intrusion detection system in internet of vehicles,'' in \emph{2019 IEEE
  global communications conference (GLOBECOM)}.\hskip 1em plus 0.5em minus
  0.4em\relax IEEE, 2019, pp. 1--6.

\bibitem{garg2020multi}
S.~Garg, K.~Kaur, S.~Batra, G.~Kaddoum, N.~Kumar, and A.~Boukerche, ``A
  multi-stage anomaly detection scheme for augmenting the security in
  iot-enabled applications,'' \emph{Future Generation Computer Systems}, vol.
  104, pp. 105--118, 2020.

\bibitem{al2019intrusion}
O.~Y. Al-Jarrah, C.~Maple, M.~Dianati, D.~Oxtoby, and A.~Mouzakitis,
  ``Intrusion detection systems for intra-vehicle networks: A review,''
  \emph{IEEE Access}, vol.~7, pp. 21\,266--21\,289, 2019.

\bibitem{SongFrequency2016}
H.~Song, H.~Kim, and H.~K. Kim, ``Intrusion detection system based on the
  analysis of time intervals of can messages for in-vehicle network,'' 01 2016,
  pp. 63--68.

\bibitem{taylor2015frequency}
A.~Taylor, N.~Japkowicz, and S.~Leblanc, ``Frequency-based anomaly detection
  for the automotive can bus,'' in \emph{2015 World Congress on Industrial
  Control Systems Security (WCICSS)}.\hskip 1em plus 0.5em minus 0.4em\relax
  IEEE, 2015, pp. 45--49.

\bibitem{seo2018gids}
E.~Seo, H.~M. Song, and H.~K. Kim, ``Gids: Gan based intrusion detection system
  for in-vehicle network,'' in \emph{2018 16th Annual Conference on Privacy,
  Security and Trust (PST)}.\hskip 1em plus 0.5em minus 0.4em\relax IEEE, 2018,
  pp. 1--6.

\bibitem{kang2016intrusion}
M.-J. Kang and J.-W. Kang, ``Intrusion detection system using deep neural
  network for in-vehicle network security,'' \emph{PloS one}, vol.~11, no.~6,
  p. e0155781, 2016.

\bibitem{hanselmann2020canet}
M.~Hanselmann, T.~Strauss, K.~Dormann, and H.~Ulmer, ``Canet: An unsupervised
  intrusion detection system for high dimensional can bus data,'' \emph{IEEE
  Access}, vol.~8, pp. 58\,194--58\,205, 2020.

\bibitem{Zhang2018ATD}
L.~Zhang, L.~Shi, N.~Kaja, and D.~Ma, ``A two-stage deep learning approach for
  can intrusion detection,'' 2018.

\bibitem{marchetti2016evaluation}
M.~Marchetti, D.~Stabili, A.~Guido, and M.~Colajanni, ``Evaluation of anomaly
  detection for in-vehicle networks through information-theoretic algorithms,''
  in \emph{2016 IEEE 2nd International Forum on Research and Technologies for
  Society and Industry Leveraging a better tomorrow (RTSI)}.\hskip 1em plus
  0.5em minus 0.4em\relax IEEE, 2016, pp. 1--6.

\bibitem{borazjani2014octane}
P.~Borazjani, C.~Everett, and D.~McCoy, ``Octane: An extensible open source car
  security testbed,'' in \emph{Proceedings of the Embedded Security in Cars
  Conference}, vol.~40, 2014.

\bibitem{malhotra2016lstm}
P.~Malhotra, A.~Ramakrishnan, G.~Anand, L.~Vig, P.~Agarwal, and G.~Shroff,
  ``Lstm-based encoder-decoder for multi-sensor anomaly detection,''
  \emph{arXiv preprint arXiv:1607.00148}, 2016.

\bibitem{marchetti2018read}
M.~Marchetti and D.~Stabili, ``Read: Reverse engineering of automotive data
  frames,'' \emph{IEEE Transactions on Information Forensics and Security},
  vol.~14, no.~4, pp. 1083--1097, 2018.

\bibitem{clevert2015fast}
D.-A. Clevert, T.~Unterthiner, and S.~Hochreiter, ``Fast and accurate deep
  network learning by exponential linear units (elus),'' \emph{arXiv preprint
  arXiv:1511.07289}, 2015.

\bibitem{kingma2014adam}
D.~P. Kingma and J.~Ba, ``Adam: A method for stochastic optimization,''
  \emph{arXiv preprint arXiv:1412.6980}, 2014.

\bibitem{sutskever2014sequence}
I.~Sutskever, O.~Vinyals, and Q.~V. Le, ``Sequence to sequence learning with
  neural networks,'' \emph{arXiv preprint arXiv:1409.3215}, 2014.

\bibitem{zago2020recan}
M.~Zago, S.~Longari, A.~Tricarico, M.~Carminati, M.~G. P{\'e}rez, G.~M.
  P{\'e}rez, and S.~Zanero, ``Recan--dataset for reverse engineering of
  controller area networks,'' \emph{Data in brief}, vol.~29, p. 105149, 2020.

\bibitem{maffiola2021goliath}
D.~Maffiola, S.~Longari, M.~Carminati, M.~Tanelli, and S.~Zanero, ``Goliath: A
  decentralized framework for data collection in intelligent transportation
  systems,'' \emph{IEEE Transactions on Intelligent Transportation Systems},
  2021.

\end{thebibliography}

\end{document}